\documentstyle[12pt,aasms4]{article}
\newcommand{\keV}{ke\kern-.15emV}
 
\begin{document}

\title{EUV and X-ray observation of Abell 2199:
a three-phase intracluster medium with a massive warm component}

\author{Richard~Lieu$\,^{1}$, Massimiliano Bonamente$\,^{1}$,
Jonathan~P.~D.~Mittaz$\,^{2}$}

\affil{\(^{\scriptstyle 1} \){Department of Physics, University of Alabama,
Huntsville, AL 35899, U.S.A.}\\
\(^{\scriptstyle 2} \){Mullard Space Science Laboratory, UCL,
Holmbury St. Mary, Dorking, Surrey, RH5 6NT, U.K.}\\
}

\begin{abstract}
Various independent ways of constraining the Hubble
constant and the baryonic content of the universe finally converged at
a consensus range of values which indicates that 
at the present epoch the bulk of
the universe's baryons is in the form of a warm
$\sim$ 10$^6$ K gas [1,2] - a temperature regime which renders them
difficult to detect.  The discovery of EUV and soft X-ray excess emission
from clusters of galaxies was originally interpreted as the first direct
evidence for the large scale presence of such a warm component [3].
We present results from an EUVE Deep Survey (DS) observation of the 
rich cluster Abell 2199 in the Lex/B (69 - 190 eV) filter passband.
The soft excess radial trend (SERT), shown by a plot 
against cluster radius $r$ of the percentage
EUV emission $\eta$ observed above
the level expected from
the hot intracluster medium (ICM), reveals that $\eta$
is a simple function of $r$ which decreases monotonically towards
$r = 0$; it smoothly  turns negative
at $r \sim$ 6 arcmin, inwards of this radius 
the EUV is absorbed by cold matter with a
line-of-sight column density of $\geq$ 2.7 $\times$ 10$^{19}$ cm$^{-2}$.  
The 
area of absorption is
much larger than that of the cooling flow. These facts
together provide strong evidence for a centrally 
concentrated but cluster-wide distribution of clumps of cold gas
which co-exist with warm gas of similar spatial properties.
Further, the simultaneous modeling of
EUV and X-ray data requires a warm component even within the region
of absorption.
The phenomenon demonstrates a three phase ICM,
with the warm
phase estimated to be $\sim$ 5-10 times
more massive than the hot.
\end{abstract}

The A2199 sky area was observed by EUVE for 
$\sim$  57 ksec in February 1999.  The
programme featured an {\it in situ} background measurement by pointing
at small offset from the cluster, which yielded an accurate
background template for point-to-point
subtraction [4].  Complementary data in
the X-ray (0.2 - 2.0 keV) passband, as gathered by
a ROSAT PSPC observation which took place in July 1990, with an exposure
of 8.5 ksec, were extracted from the public archive [5].
For correct comparison between the EUV and X-ray emissions, the
Galactic HI column density was measured at
N$_H$ = (8.3 $\pm$ 1.0) $\times$ 10$^{19}$ cm$^{-2}$
by a dedicated observation
at Green Bank [6], and was found to be spatially smooth.
The EUV and X-ray data were simultaneously modeled with a
thin plasma emission code [7,8] and appropriate line-of-sight Galactic
absorption [9] for the above value of N$_H$.  At a given radius the hot ICM
was assumed to be isothermal, with the abundance fixed at
0.3 solar apart from the cooling flow region where the parameter
became part of the data fitting in order to account for any possible
abundance gradient within this region (a different way of handling
the abundance does not sensitively affect the results presented in
this work).

It is found that the forementioned model, when applied to the
PSPC spectra at all radii, generally leads to acceptable fits.  
At low energies the EUV measurements gave crucial
new information.  The overall effect is a soft excess as
reported previously [4].  A plot of the SERT indicates, however,
that the average percentage EUV excess at a given radius is
less at the centre, see Figure 1.  In fact, the trend takes
the form of a very negative central excess (i.e. absorption), which
steadily rises with radius until the
6 arcmin point, beyond which this fractional excess turns positive and
continues to increase until the limiting radius of EUV
detection ($\sim$ 20 arcmin [4]).

We first address the outer parts of the cluster, where the data
already demonstrated the implausibility of a non-thermal interpretation
of the soft excess (which postulates a large population of
intracluster relativistic electrons undergoing inverse-Compton (IC)
interaction with the cosmic microwave background (CMB)
[10,11]).  
Figure 2 shows a composite plot of 
the EUV and X-ray data for the 12 - 15 arcmin annulus.  The
prominent EUV excess, unaccompanied by any similar effect in
soft X-rays, implies that the bulk of the relativistic  electrons
have energies below 200 MeV, a cut-off which is most
obviously understood as due to aging (i.e. synchrotron and
inverse-Compton losses): the electrons are at
least 3 $\times$ 10$^9$ years old.  However, in order to account 
for the large EUV excess the highly evolved 
electron spectrum at the present epoch must still include sufficient
particles ahead of the cut-off.  This means that for the region
of concern, at
injection (when the power-law differential number index is
assumed to be 2.5, in accordance with our Galactic cosmic ray
index) the relativistic electron pressure
would have
exceeded that of the hot ICM by a factor of $\sim$ 4, leading to
a major confinement problem for the hot gas.  The inclusion of
cosmic ray protons exaggerates the difficulty, as
protons carry 10 - 100 times more pressure than electrons.
Thus, by elimination, the only viable alternative, viz. the
originally proposed
thermal (warm) gas scenario [3], must now be considered seriously.
This is especially so in the light of the recent constraints
on cosmological parameters, which point to the existence of
a warm and massive baryonic component, as mentioned earlier.

To appreciate the multi-phase nature of the ICM of A2199,
we move radially inwards, where Figure 1
indicates that the EUV is absorbed.  For
more details, we show in Figure 3 an image of the 
of the fractional EUV
excess $\eta$.
The data suggest an intermixed model [12]
of the ICM: the lack of soft excess 
at small radii is due entirely to
the larger amount of cold absorbing matter collected in this region.  
Thus, while in the north-south direction
severe absorption persists
out to a radius of $\sim$ 4 arcmin, in the east-west direction
signatures of soft excess are already present as close as $r \sim$ 2 arcmin.

Our inference of the state of the ICM is reinforced by the behavior of the
SERT: it follows a simple parametric
profile which applies equally satisfactorily to the absorption
and soft excess regions, 
with no change of behavior at the transition radius of $\sim$ 6 arcmin
(Figure 1).
In fact, there is no particular significance in this radius (it is
much larger than the cooling flow radius of $\sim$ 2 arcmin [13]),
The observation is naturally interpreted as the combined effect of 
clumped emission regions containing a warm
component, absorbed by blobs of cold gas sandwiched in between.  
Both distributions are
cluster-wide and centrally condensed, but with increasing radius 
the lines-of-sight are more
transparent to EUV photons created at locations along them.
For comparable intrinsic emission profiles of the soft
excess and the hot ICM, the
result is an outwardly rising SERT.  

The correctness of this
approach is confirmed by our study of the nearby Virgo and Coma
clusters.  In the former case a strong SERT exists despite no
apparent absorption (i.e. $\eta > 0$ everywhere).  Yet a spatial
analysis of the PSPC image 
showed that a central circular area has a statistically
significant enhancement in the
number of small regions where the soft excess brightness
is below the mean value for this circle
(i.e. some of them must be due to resolved
absorption clouds); with the effect disappearing gradually
towards annuli of larger radius.  
For Coma the SERT is very weak, implying little absorption, and indeed
a similar analysis revealed no evidence of clouds at any radius.
Results on these two clusters will be published shortly.

The argument for an intermixed ICM also rests upon
direct evidence for the presence of
soft excess even in the absorbed regions.  We show in Figure 4
a core spectrum, where it can be seen that by the time intrinsic
absorption accounts for the
EUV decrement, an excess is seen in soft X-rays
(which are less absorbed).  This clearly
indicates a complex ICM where the various gas phases co-exist.  The
apparent negative soft excess within the absorption radius is
simply due to an abundance of cold clouds masking EUV emissions from
the warm and hot components.

The thermal origin of the EUV is compelling for another reason:
the widespread absorption reported here implies the existence of 
a cold phase in the midst of the well known hot phase, and the question
then naturally arises concerning why a warm phase is absent, and
is not the cause of the soft excess.  At the very least, mixing
layers on the surface of the cold clouds would suffice to produce
the intermediate phase [14].

The mass budgets of the three ICM components in consideration
are estimated as follows.  The intrinsic HI column density as
inferred from the central EUV absorption converts to
a density of cold clouds of $\sim$ 5 $\times$ 10$^{10}$
M$_\odot$ Mpc$^{-3}$.
This gives a mass ratio of 1:2000 between
the cold and hot gas along the line-of-sight.  Any estimate of
the mass of warm gas at the centre is likely to be inaccurate,
since the soft emission is significantly absorbed.  We therefore
considered the 12 - 15 arcmin region where this complication is
not as severe as in the center.
The extreme softness of the
emission (Figure 2) limits the gas temperature to kT $<$ 100 eV
(or T $<$ 10$^6$ K), with a correspondingly large mass estimate
of 1.25 $\pm^{0.4}_{0.9} \times$ 10$^{14}$ M$_\odot$, 
i.e. $\sim$ 43 $\pm^{13}_{29}$ times more massive than the hot ICM
in this region.  The 1-$\sigma$ lower 
limit ratio implies $\sim$ 3 times more missing
baryons than expected [1], although it must be emphasized that
both the mass and bolometric luminosity can be substantially
reduced if the gas turns out to be warmer.  This can be realized
by adopting alternative emission models for the warm phase,
especially those which involve
an underionized gas, since the EUV emission efficiency is then
enhanced, and the gas can be warmer than the above temperature constraint.
Plasma in such an ionization state is easily produced by mixing
layers or shock heating.  Another possibility is that the gas
is actually warmer than our inferred temperature, 
and the lack of a soft X-ray excess is only an
apparent effect caused by
residual absorption at these outer radii.

\vspace{3mm}

\noindent
{\bf Figure Captions}

\noindent
Figure 1:  The SERT effect illustrated by a plot against cluster
radius $r$ of the EUV fractional excess $\eta$, defined as $\eta = (p - q)/q$,
where $p$ is the DS Lex/B observed signal and $q$ is the expected
EUV emission from the hot ICM.  $q$ is determined from the best model of
the PSPC data (single temperature fits were found to be
satisfactory at all radii) with details of Galactic absorption
as quoted in the text.  The data follow a parametric profile
$\eta = -0.45+0.0075 \; r^{2.5}$ (solid line).

\noindent
Figure 2: Emission models (solid line)
used to simultaneously fit the EUVE/DS
and ROSAT/PSPC data of the 12 - 15 arcmin annular region of A2199.
{\it Left Panel:} isothermal thin plasma spectrum [16-18]
at kT = 4.08 keV [19] and an abundance of 0.5 solar.  Note
the strong EUV excess recorded by the DS (left most data point)
which is not seen in soft X-rays by the PSPC (remaining data points).
{\it Right Panel:} same as the previous model, except with an
additional non-thermal component due to the IC/CMB effect (see text).
The electron population (assumed
to have an initial differential number index of 2.5, similar to
that of Galactic cosmic rays) is
$\sim$ 3.5 Gyr old, as during this period the IC/MWB and
synchrotron losses would have secured the necessary high energy
cut-off to avoid emissions in the PSPC
passband.  At the present
epoch the electron pressure is $\sim$ 25 \% that of the
hot ICM, while the initial value of this ratio was $\sim$ 400 \%.

\noindent
Figure 3: An image of the surface brightness of EUV excess for
the central region of Abell 2199, obtained after subtraction
of background and contributions from the hot ICM emission (see
text).  The pixel units (color coded) are in 10$^{-3}$ 
photons arcmin$^{-2}$ s$^{-1}$. 
Pixels of negative excess correspond to areas where the
EUV from warm and hot components are absorbed by a cold
component.  The common centroid of the cluster EUV and soft X-ray
emissions is marked by a cross.

\noindent
Figure 4: Data are as in Figure 2, except for the 1 -- 2 arcmin
radius of A2199.  {\it Left Panel:} single temperature
emission model (kT = 3.58 $\pm^{1.07}_{0.69}$,
 abundance = 0.56$\pm^{0.43}_{0.25}$ solar) showing
the EUV signal in absorption.  {\it Right Panel:} Plasma properties
as above, with an intrinsic cold gas of line-of-sight HI column density 
$N_H$ = 2.7 $\times$ 10$^{19}$
cm$^{-2}$ invoked to account for the depleted EUV flux.  Note this
correction reveals
a soft X-ray excess in the PSPC 1/4- keV band, thus clearly indicating
the presence of an underlying warm component which is masked by
the cold absorbing phase.

\vspace{3mm}

\noindent
{\bf Referenceseferences}

\noindent
1. Cen, R. and Ostriker, J.P. 1999, {\it ApJ}, {\bf 514}, 1-6. \\
\noindent
2. Maloney, P.R. and Bland-Hawthorn, J. 1999, {\it ApJ}, {\bf 522}, L81-84. \\
\noindent
3. Lieu, R., Mittaz, J.P.D., Bowyer, S., Breen, J.O.,
Lockman, F.J., \\
\indent Murphy, E.M. \& Hwang, C. -Y. 1996b, {\it Science}, {\bf 274},1335--1338. \\
\noindent
4. Lieu, R., Bonamente,M. ,Mittaz, J.P.D., Durret, F., Dos Santos, S. and \\
\indent Kaastra, J.S.  1999, {\it ApJ}, {\bf 527}, L77 \\
\noindent
5. See the High Energy Astrophysics Archive available at \\
\indent http://heasarc.gsfc.nasa.gov/docs/rosat/archive.html . \\
\noindent
~6. Kaastra, J.S., Lieu, R.

\noindent
1. Cen, R. and Ostriker, J.P. 1999, {\it ApJ}, {\bf 514}, 1-6. \\
\noindent
2. Maloney, P.R. and Bland-Hawthorn, J. 1999, {\it ApJ}, {\bf 522}, L81-84. \\
\noindent
3. Lieu, R., Mittaz, J.P.D., Bowyer, S., Breen, J.O.,
Lockman, F.J., \\
\indent Murphy, E.M. \& Hwang, C. -Y. 1996b, {\it Science}, {\bf 274},1335--1338. \\
indent
4. Lieu, R., Bonamente,M. ,Mittaz, J.P.D., Durret, F., Dos Santos, S. and \\
\indent Kaastra, J.S.  1999, {\it ApJ}, {\bf 527}, L77 \\
\noindent
5. See the High Energy Astrophysics Archive available at \\
\indent http://heasarc.gsfc.nasa.gov/docs/rosat/archive.html . \\
\noindent
~6. Kaastra, J.S., Lieu, R.
 Mittaz, J.P.D.,
 Bleeker, J.A.M., Mewe, R., 
 Colafrancesco, S. and \\
\indent Lockman, F.J. 1999, {\it ApJ}, {\bf 519}, L119--122.\\
\noindent
~7. Mewe, R., Gronenschild, E.H.B.M. and van den Oord, G.H.J. 1985, {\it A \& A}, {\bf 62}, 197 .\\
\noindent
~8. Mewe, R., Lemen, J.R., and van den Oord, G.H.J. 1986, {\it A \& A}, {\bf 65}, 511 .\\
\noindent
~9. Morrison, R. and McCammon, D. 1983, {\it ApJ}, {\bf 270}, 119.\\
\noindent
~10. Sarazin, C.L., Lieu, R. 1998, {\it Astrophys. J.}, {\bf 494}, L177--180. \\
\noindent 
~11. Ensslin, T.A., Biermann, P.L. 1998, {\it Astron. Astrophys.}, {\bf 330},
90--98. \\
\noindent
~12. Jakobsen, P. and Kahn, S.M. 1986, {\it ApJ}, {\bf 309}, 682 .\\
\noindent
~13. Siddiqui, H., Stewart, G.C. and Johnston, R.M. 1998, 
{\it Astr. Astrophys.}, {\bf 334}, 71--86.\\
\noindent
~14. Fabian, A.C. 1997, {\it Science}, {\bf 275}, 48--49. \\

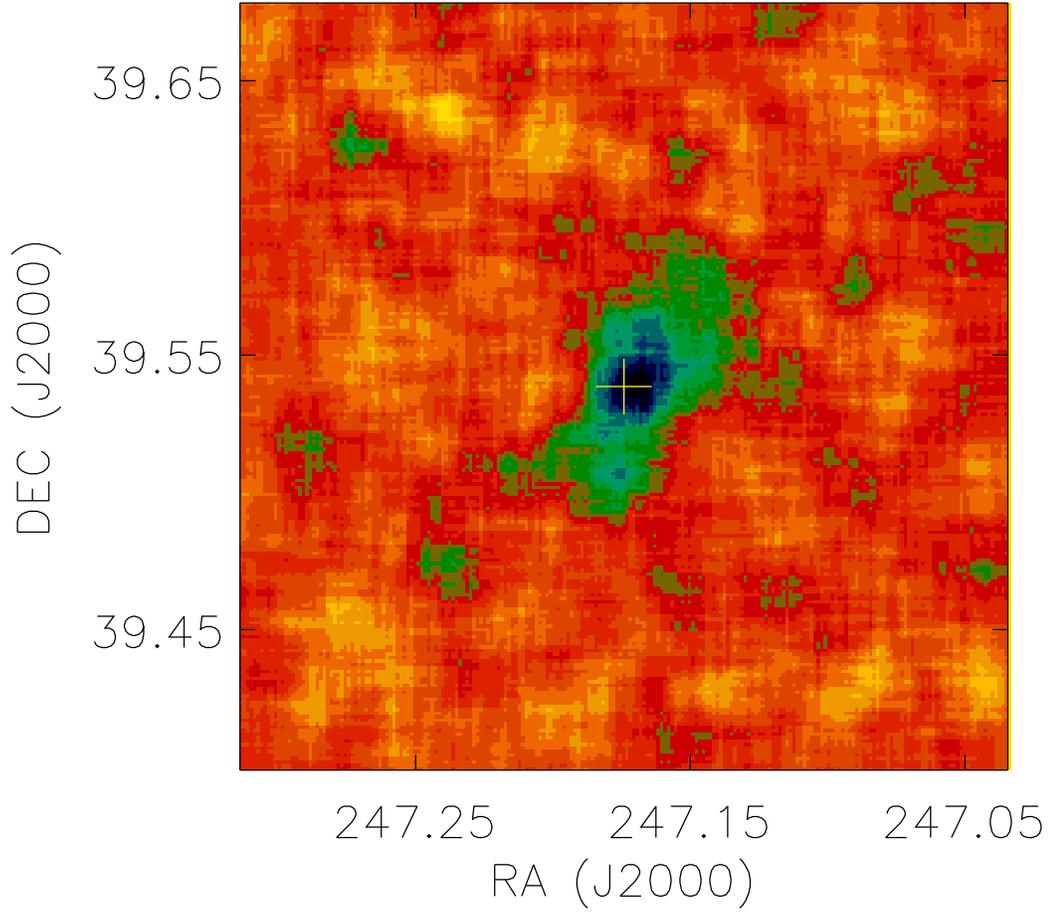
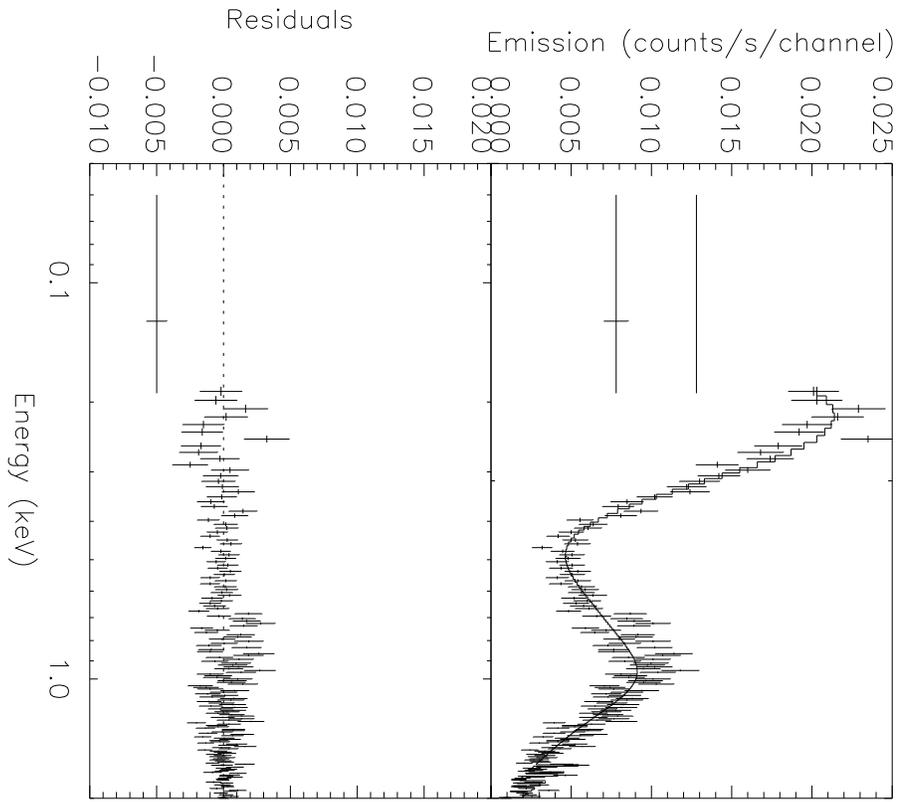
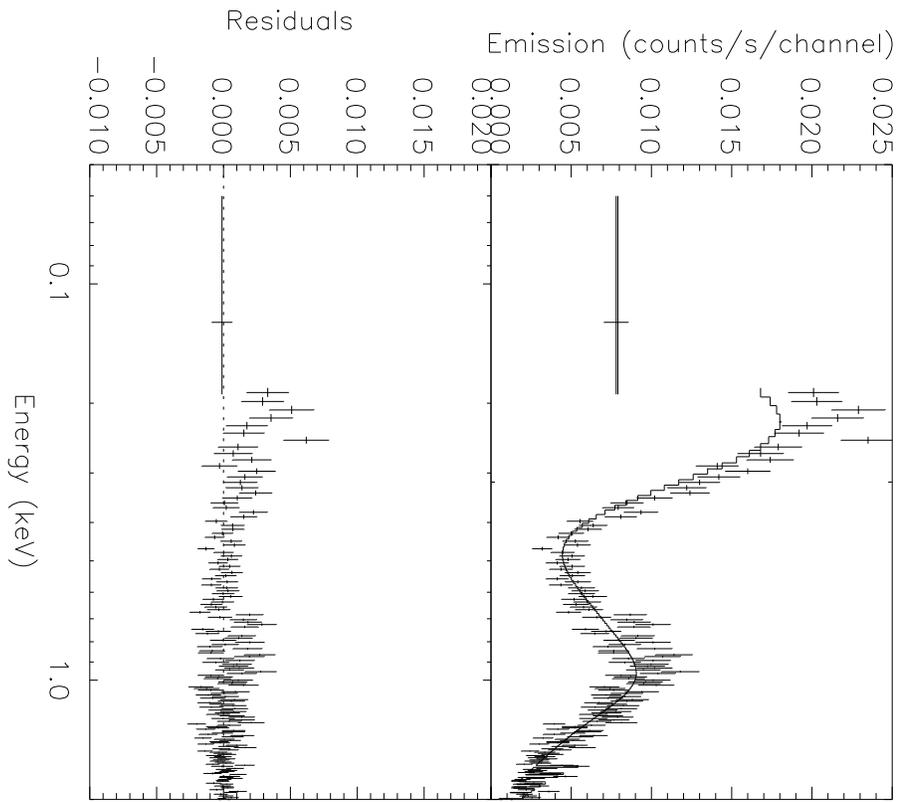
\end{document}